\documentclass[12pt,a4paper]{article}
\usepackage{epsfig}
\usepackage[hang,nooneline]{subfigure}
\usepackage{amsmath}
\usepackage{amsfonts}
\usepackage{amssymb}
\usepackage{graphicx}
\usepackage{epstopdf}
\usepackage{amsmath}
\usepackage{graphicx} \usepackage{float}
\usepackage{amsthm}
\usepackage{comment}
\usepackage{color}
\usepackage{graphicx}
\usepackage{amscd}
\usepackage{hyperref}
\usepackage{listings}
\usepackage{framed}
\usepackage[backend=biber,style=numeric]{biblatex}
\lstnewenvironment{mat}
{\lstset{language=mathematica,mathescape,columns=flexible}}
{}

\definecolor{shadecolor}{gray}{0.9}

\def\mathcolour#1#{\mathcoloraux{#1}}
\newcommand*{\mathcoloraux}[3]{%
  \protect\leavevmode
  \begingroup
    \color#1{#2}#3%
  \endgroup
}

\usepackage[svgnames]{xcolor}

\usepackage{xcolor}

\definecolor{blue}{HTML}{1F77B4}
\definecolor{orange}{HTML}{FF7F0E}
\definecolor{green}{HTML}{2CA02C}

\begin{document}
\title{\vspace{-2cm}Algorithmic Information Dynamics\\ of Cellular Automata\thanks{Invited contribution to The Mathematical Artist: A Tribute to John Horton Conway, by WSPC.}}

\author{\vspace{0.5cm}Hector Zenil$^{1,2,3,4}$ and Alyssa Adams$^{4,5}$\\ 
$^1$ Oxford Immune Algorithmics, Reading, RG30 1EU, U.K.\\
$^2$ Alan Turing Institute, British Library, London, NW1 2DB, U.K.\\
$^3$ Algorithmic Dynamics Lab, Karolinska Institute, Stockholm,\\ 171 77, Sweden\\
$^4$ Algorithmic Nature Group, LABORES, Paris,\\76006, France\\
$^5$ Morgridge Institute of Research \& Department of\\Bacteriology,
University of Wisconsin-Madison,\\Madison, WI, 53706, U.S.A.}

\date{}

\maketitle              

\begin{abstract}
We illustrate an application of Algorithmic Information Dynamics (AID) to Cellular Automata (CA) demonstrating how this digital calculus is able to quantify change in discrete dynamical systems.  We demonstrate the sensitivity of the Block Decomposition Method on 1D and 2D CA, including Conway's Game of Life, against measures of statistical nature such as compression (such as Lempel–Ziv–Welch) and Shannon Entropy in two different contexts (1) perturbation analysis and (2) dynamic-state colliding CA. The approach is interesting because it analyses a quintessential object native to software space (CA) in software space itself by using algorithmic information dynamics through a model-driven universal search instead of a traditional statistical approach e.g. LZW compression or Shannon entropy.  The colliding example of two state-independent (if not three as one is regulating the collision itself) discrete dynamical systems offers a potential proof of concept for the development of a multivariate version of the AID calculus.\\

\end{abstract}

\section{Introduction}

In the Summer of 2008, John H. Conway and I (HZ) had the chance to meet each other as instructors of different Summer Schools held in the same place, the University of Vermont at Burlington, U.S. One of the main interests of John was the study of rich dynamics of very simple discrete systems such as cellular automata. We lost John to complications of COVID-19, a disease I am now trying to fight using the same kind of mathematical tools he was interested in that I call Algorithmic Information Dynamics.

Algorithmic Information Dynamics (AID)~\cite{Zenil:2020} is an algorithmic probabilistic framework for causal discovery and causal analysis. It enables a numerical solution to inverse problems based or motivated on principles of algorithmic probability. AID studies dynamical systems in software space where all possible computable models can be found or approximated under the assumption that discrete longitudinal data such as particle orbits in state and phase space can approximate continuous systems by Turing-computable means. AID combines perturbation analysis and algorithmic information theory to guide a search for sets of models compatible with observations and to precompute and exploit those models as testable generative mechanisms and causal first principles underlying data and systems. AID is an alternative or a complement to other approaches and methods of experimental inference, such as statistical machine learning and classical information theory.

One may ask how relevant a purely theoretical framework based on computable discrete models can be to the real world, but it is never too difficult to find simple arguments allowing algorithmic explanations to complex phenomena~\cite{zenillife,Zenil2019,zenilgeo}. AID connects with and across other parallel fields of active research such as logical inference, causal reasoning, and symbolic computation. AID studies how candidate discrete computable equations as generating mechanisms are affected by changes in observed phenomena over time as a result of a system evolving (e.g. under the influence of noise) or being externally perturbed.

Unlike graphical methods such as Bayesian networks, AID does not rely on graphical representations or (often inaccessible) empirical estimations of mass probability distributions. AID encompasses the foundations and methods that make the area of algorithmic information and algorithmic complexity more relevant to scientific discovery and causal analysis.

We provide an overview of algorithmic information dynamics and illustrate its methods in application to elementary cellular automata.

\section{Cellular Automata}

A {\it cellular automaton} (CA) is a tuple $\langle S, (\mathbb{L}, +), T, f \rangle$  with a set $S$ of states, a lattice $\mathbb{L}$ with a binary operation $+$, a neighbourhood template $T$, and a local rule $f$.

The {\it set of states} $S$ is a finite set with elements $s$ taken from a finite alphabet $\Sigma$ with at least two elements. It is common to take an alphabet composed entirely of integers modulo $s$: $\Sigma = \mathbb{Z}_s = \{0,...,s-1\}$. 
An element of the lattice $i\in \mathbb{L}$ is called a cell. The lattice $\mathbb{L}$ can have $D$ dimensions and can be either infinite or finite with cyclic boundary conditions.

The {\it neighbourhood template} $T=\langle\eta_1,...,\eta_m\rangle$ is a sequence of $\mathbb{L}$. In particular, the neighbourhood of cell $i$ is given by adding the cell $i$ to each element of the template $T$: $T=\langle i+\eta_1,...,i+\eta_m\rangle$. 
Each cell $i$ of the CA is in a particular state $c[i] \in S$. A {\it configuration} of the CA is a function $c: \mathbb{L} \rightarrow S$. The {\it set of all possible configurations} of the CA is defined as $S_\mathbb{L}$.

The {\it evolution of the CA} occurs in discrete time steps $t=0,1,2,...,n$. The transition from a configuration $c_t$ at time $t$ to the configuration $c_{(t+1)}$ at time $t+1$ is induced by applying the local rule $f$. The local rule is to be taken as a function $f: S^{|T|} \rightarrow S$ which maps the states of the neighbourhood cells of time step $t$ in the neighbourhood template $T$ to cell states of the configuration at time step $t+1$:
\begin{equation}
c_{t+1}[i]=f\left(c_t[i+\eta_1],...,c_t[i+\eta_m]\right )
\end{equation}
The general transition from configuration to configuration is called the {\it global map} and is defined as: $F: S^\mathbb{L} \rightarrow S^\mathbb{L}$.

In the following we will consider 1-dimensional (1-D) CA as introduced by Wolfram~\cite{StephenWolfram1983,nks}. The lattice can be either finite, i.e. $\mathbb{Z}_N$, having the length $N$, or infinite, $\mathbb{Z}$. In the 1-D case it is common to introduce the {\it radius} of the neighbourhood template which can be written as $\langle -r,-r+1,\dots ,r-1,r \rangle$ and has length $2 r+1$ cells. With a given radius $r$ the local rule is a function $f: \mathbb{Z}_{|S|}^{{|S|}^{(2r+1)}} \rightarrow \mathbb{Z}_{|S|}$ with $\mathbb{Z}_{|S|}^{{|S|}^{(2r+1)}}$ rules. The so called Elementary Cellular Automata (ECA) with radius $r=1$ have the neighbourhood template $\langle -1,0,1\rangle$, meaning that their neighbourhoods comprise a central cell, one cell to the left of it and one to the right. The rulespace for ECA contains $2^{2^{3}}=256$ rules. 

Two-dimensional cellular automata were studied by John Conway~\cite{conway} of which his Game Of Life (GoL) is its most popular example.  A more comprehensive analysis of the dynamics of GoL and its persistent particles is provided in~\cite{zenilgol}



Wolfram introduced~\cite{nks} an heuristic for classifying computer programs by inspecting the behaviour of their space-time diagrams. Computer programs behave differently for different inputs. It is possible, and not uncommon, however, to analyze the behaviour of a program asymptotically according to an initial condition metric~\cite{zenilchaos}.

\section{Algorithmic Information Dynamics (AID)}

Based upon or motivated by algorithmic probability in its modern formulation as introduced by Kolmogorov, Chaitin, Solomonoff and Levin ~\cite{Chaitin1987-kt, Calude2002-da, Downey2010-tq, Li2019-op}, the field of Algorithmic Probability (AP) considers the probability of a (discrete) object being produced by an algorithm runing on a Turing universal system. AP imposes a mapping distribution called the universal distribution between input and output. Formally, a computable process that produces a string s is a program p that when executed on a universal Turing machine $U$ produces the string s as output and halts.

As $p$ is itself a binary string, we can define the discrete universal a priori probability $m(s)$ as the probability that the output of an arbitrary binary input of a universal prefix-free Turing machine $U$ is s when this input is provided with fair coin flips on the input tape. Formally,

\begin{equation}
m(s):=\sum_{p:U(p)=s}2^{-l(p)}
\end{equation}

where the sum is over all halting programs $p$ for which U outputs the string $s$. As $U$ is a prefix-free universal Turing machine, the set of valid programs forms a prefix-free set (or self-delimited programming language), and thus the sum is bounded, given Kraft's inequality (i.e., it defines a probability semi-measure).

The methods underpinning AID can be described as a combination of Bayes' theorem and computability theory, where the default agnostic prior distribution is the universal distribution instead of some other agnostic distribution such as the uniform distribution. In this way, any updates to the specific distribution of the problem proceed according to algorithmic probability.

Introduced by Ray Solomonoff and Leonid Levin~\cite{Solomonoff1964-hm,Levin1974-ci}, algorithmic probability, or the theory of universal inductive inference as he also called it, is a theory of prediction based on observations that assumes that each individual observed object is generated by arbitrary computable processes. An example is the prediction of the next digit in the sequence $s=1234567\ldots .$ According to algorithmic probability, the next digit would be 8; if the simplest model (i.e., the shortest self-delimiting program) able to generate that sequence is the successor function $x_{0}=1;x_{i}=x_{i-1}+1,$ such a model would generate 8 as a predictor of the next digit given the data. The only assumption is that the prior probability follows a computable probability distribution even if such a distribution is unknown because as proven by Levin, all computable distributions converge in what is known as the universal distribution. As one of the pillars of algorithmic information theory (AIT), the universal distribution that is an offshoot of the algorithmic coding theorem conjoins algorithmic complexity, universal a priori probability, and a universal/maximal computably enumerable discrete semi-measure into a single precise and ubiquitous mathematical formalization of a necessary ``bias toward simplicity" for spaces of computably generated objects. This is a result that has been called 'miraculous' in the scientific literature and been lauded by the late Marvin Minsky as the most important scientific theory of relevance to AI~\cite{Zenil2020-xw, Zenil2019-tk}.

Algorithmic probability captures (albeit without actually setting out to do so) longstanding principles on which science has been founded: the principle (also known as Occam's razor) that the simplest explanation (in this case the most algorithmically probable which turns out to be the shortest algorithmic description) is most likely the correct one; and the principle of multiple explanations (Epicurus), which mandates the retention of all explanations consistent with the data; and Bayes's Rule, requiring the transformation of the a priori distribution into a posterior distribution according to the evidence, to keep all hypotheses consistent with the data~\cite{Zenil2020-xw}.

\subsection{Numerical Methods}

AID was conceived and introduced in the early 2010s and is currently an active area of research, but the methods enabling AID were introduced in the mid-2000s with the first publication of a calculation and systematic study of output probability distributions of different types of models of computation appeared in (Delahaye \& Zenil, 2007). While it was not known--no experimental verification being then available--whether the concept of algorithmic probability would empirically capture the intuition of Occam's razor other than as established in the theory, Zenil and colleagues studied the behavior of these probability distributions in an exhaustive fashion ~\cite{Delahaye2007-oz, Soler-Toscano2014-er}. Their results suggested that the output distributions were more compatible than theoretically expected, and furthermore, met both the theoretical and empirical expectations entailed in Occam's razor, as the elements assigned the highest probability were also found to be the most algorithmically simple according to various complementary order parameters, including computable ones, which converge in value and thus provide further confirmation, while the least frequent elements were also more random by all measures ~\cite{Zenil2020-xw, Zenil2020a}.

The numerical application of AID beyond its theoretical formulation relies strongly upon numerical methods designed to encompass and expand classical information theory to characterize randomness, using measures other than popular compression algorithms, such as Lempel-Ziv, widely used to produce purported approximations to algorithmic complexity~\cite{Zenil2020-xw} . These methods on which AID relies upon (but is also independent of) are the so-called coding theorem (CTM) and block decomposition (BDM) methods that have as their main features (1) that can go beyond entropic and statistical compression approaches by providing the means to explore and find computable models allowing causal discovery (and as opposed to e.g. obfuscated compressed files with no state correspondence to a model or evolving dynamical system), and (2) are sensitive enough to allow (small) perturbation causal analysis~\cite{zenil_2018, Zenil2019, zenil_2019b}.

\subsection{The Coding Theorem Method and Causal Discovery}

Most attempts to approximate algorithmic complexity have proceeded by way of popular lossless compression algorithms such as LZ or LZW. However, if we take the example of the sequence $s=1234567\ldots$, an algorithm such as LZ or LZW will fail at compressing s despite its obvious compressed form $(x_{0}=1;x_{i}=x_{i-1}+1)$. This is because algorithms such as LZ and LZW are entropy estimators and they build a dictionary of most frequent words to assign them shorter codes. For example, if $s$ is a computable Borel normal number ~\cite{Calude2002-da}, no contiguous subsequence is approximately represented more frequently than any other of the same length, and the resulting compressed file tends to be of about the same size as s itself (modulo change of data type transformation, which is only a transliteration from, e.g., ASCII to binary) ~\cite{Zenil2017a}. Because of these limitations of popular lossless compression algorithms, and for other reasons, Zenil et al. introduced a method based on the so-called algorithmic coding theorem. Formulated by L.A. Levin ~\cite{Levin1974-ci}, the algorithmic coding theorem in the context of algorithmic probability establishes equality between universal a priori probability $m(s)$ and algorithmic complexity $K(s)$. Formally:

\begin{equation}
m(s)=2^{-K(s)}+c
\end{equation}

or equivalently,

\begin{equation}
-log\:m(s)=K(s)+c,
\end{equation}
where c is a constant.

Based on this fundamental theorem, and under the assumption of optimality of the reference Turing machine, the Coding Theorem Method (CTM) is an alternative to statistical compression algorithms such as Lempel-Ziv (LZW)~\cite{zenilreview}, widely used to approximate algorithmic complexity. CTM does not rely upon statistical methods such as those based on the dictionaries on which popular lossless compression algorithms such as LZW are based (being designed to find statistical regularities and thus more closely related to classical information theory than to algorithmic complexity)~\cite{Delahaye2012-eb, Soler-Toscano2014-bj, Soler-Toscano2013-fm, Zenil2015-dx}. The aim of CTM is to embrace Turing universality and explore the space of computer programs able to capture properties beyond statistical patterns (as a consequence of which has been overlooked in statistical approaches). Then, the method pumps it into the generative models that support a natural or artificial phenomenon, following the usual scientific modus operandi--whereas in statistical and probabilistic approaches the part of the probabilistic content of an object or state that belongs to the model itself and the part that belongs to the explanation of the possible explanatory models have traditionally been conflated.

For example, when modeling the outcome of throwing a dice, what a statistical model ends up quantifying is an uncertainty extrinsic to the process itself, the degree of uncertainty of an observer and its inability to determine the underlying nature of the generating mechanism (if any). Thus, it quantifies a property of the observer and not the process undergone by the dice itself. This is because the process of throwing the dice is deterministic according to the laws of classical mechanics, which are known to govern the dice trajectory (though quantum fluctuations may be believed to be probabilistic, at short distances they wouldn't have any effect). The probabilistic content of the model describing the dice outcome is thus extrinsic to the actual process. What an algorithmic-probability-like approach like CTM would do in principle--and this is its major difference--is to produce a set of deterministic models describing the dice trajectory and outcome without recourse to probability. Consequentially, it is in the distribution of possible computable models explaining the dice that a probability emerges; it is not assumed by the individual models themselves. In turn, not only can these models be tested beyond their outcome predictions as mechanistic (step-by-step) descriptions of the process, but they also offer a non-black-box approach where each model can be followed step-by-step, with model states corresponding to constructive (e.g. physical) states, as opposed to random variables with no state-to-state correspondence between the model and phenomenological data.

Unlike popular lossless compression algorithms that are guaranteed not to be able to characterize objects such as $s$, CTM can in principle do so, because when running all possible computer programs up to the size of $s$ in bits there is a non-zero probability that CTM will find such a program if it exists. Indeed, this is guaranteed, as the worst case is when the program in question has the form of print[s] and s is algorithmically random (i.e., $s$ is incompressible or with an algorithmic complexity value equal to the length of s, up to a constant). This holds because of the ergodicity of the algorithmic complexity approximation by the CTM over the software space, if enough computational resources are expended, which in turn is enabled by the lower semi-computability of the universal a priori probability ~\cite{zenil_2018, Delahaye2012-eb}.

Other measures such as Granger causality and transfer causality are methods of statistical in nature and used mostly for causal analysis, not causal discovery, especially because they are unable to deal with models in phase-space corresponding to possible physical state models. Some relationships between variables, causal or temporal, can be derived as they can be from intervention analyses similar to Pearl's do-calculus, but they don't produce candidate mechanistic models in the first place, and the causal analysis falls back into the purely classical probabilistic framework which, though partially circumvented, must be resorted to again when it comes to testing the associative nature of 2 or more variables.

\subsection{The Block Decomposition Method (BDM)}

One way to see BDM is as a weighted version of Shannon's entropy that introduces a local quantification of algorithmic complexity into the original formulation of classical information, a version that is able to help tell apart statistical randomness and algorithmic randomness~\cite{zenil_2018, Zenil2019, Zenil2020-xw}.  This is a key epistemological distinction because (1) a sequence such as $s=123456\ldots$ would be characterised as maximally `disordered' by statistical approaches such as Shannon Entropy for an observer unaware of its deterministic nature (without concept of natural number) as it is a Borel normal number (Champernowne), unless upon its application one already had the information that $s$ had been generated by a deterministic process, which renders the use of the measure redundant (in science this is the rule rather than the exception when observing data: one investigates the nature of an object when the nature of that object is as yet unknown); and (2) it is of high empirical value in application to science.

For example, if we set out to quantify human memory, an experimental subject would not need to learn s digit by digit in order to generate it, illustrating the algorithmic nature of human cognitive processes, that go beyond statistical patterns. The sequence s may look very special, but in fact, most sequences are of this type. In addition to not having any statistical regularity, most sequences--if we consider the set of all possible sequences--do not even have any short description (low algorithmic randomness). That is, most sequences are algorithmically random. Moreover, the difference between those that do not have a short algorithmic description versus those that appear statistically random but have a short description is divergent. We only chose $s$ because it was the most obvious for purposes of illustration (another example would be the digits of a mathematical constant such as $\pi$). Indeed, CTM and BDM have found many applications in psychometrics and cognition due to this advantage (see Applications of AID).

What BDM does is to extend the power of CTM to quantify algorithmic randomness by implementing a divide-and-conquer approach whereby the data is decomposed into pieces small enough that an exhaustive computational search finds the set of all computable models able to generate each piece of data~\cite{zenil_2018, Zenil2019, Zenil2020-xw}. The sequence of small generators then supports the larger piece of data from which insight is gained as to their algorithmic properties. Small pieces of data supported by even smaller programs than those pieces are called causal patches in the context of AID, as they are the result of a causal relationship between the set of these computable models and the observed data. On the other hand, models of about the same length as the data are not causally explained by a shorter generating mechanism and are therefore considered (algorithmically) random and not causally supported~\cite{Zenil2020a}.

A sequence of computer programs smaller than their matched patches constitutes a sufficient test for non-randomness and is therefore informative of its causal content, as its components can be explained by underlying small computable models shorter than the original whole data itself. This way, the stronger the coarse-graining of the original data (i.e., the weaker the decomposition) the closer the value resulting from the BDM is to the theoretical optimal value that corresponds to the whole data's algorithmic complexity. Hence, the more fine-grained (i.e., the stronger the decomposition), the more the BDM value can diverge from the whole data's algorithmic complexity. This is because the BDM value and the algorithmic complexity are proven to converge (up to a constant that only depends on the error of the CTM relative to the algorithmic complexity) as the data partition size tends to the whole data size ~\cite{zenil_2018}.

\subsection{Algorithmic Intervention Analysis}

AID, as based on CTM, is a resource-bounded algorithmic complexity measure, as it takes informed runtime cutoffs (usually very short, to deal with short strings) to estimate candidate upper bounds of algorithmic complexity under assumptions of optimality and for the reference universal enumeration chosen.

While AID, MML, and MDL are ultimately related and are based on the principles of algorithmic probability, these minimum length approaches depart from AID (and AIT) in fundamental ways. MDL is a model selection principle rather than a model generating method. As such, AID incorporates the principle of MDL and represents a generalization, with MDL being a particular case based on learning algorithms using the statistical notion of information rather than the more general--and powerful--notion of algorithmic information used by AID that requires Turing-completeness. In practice, however, MDL and AID can complement each other because AID is more difficult to estimate while MDL provides some statistical shortcuts, and this is similar to and mostly already exemplified in the Block Decomposition Method (BDM), which combines both classical and algorithmic information theories.

In fact, AID is agnostic as regards the underlying method, even if it currently relies completely on the use of CTM and BDM, as described above. So, for example, when AID confines itself to statistical models not produced by CTM, it would collapse into methods such as MDL or MML, but any methods that introduce attempts to go beyond the statistical would approximate the spirit (if not the measure, under the assumption of optimality) of AID as exemplified in CTM and BDM.

AID can be considered a special case or generalization of the area of computational mechanics, depending on one's perspective. They differ in that (1) Crutchfield's formulation involves the introduction of stochastic processes into the models themselves (rather than on the probability distribution of the models), while with AID there is no probabilistic content at the core of the candidate generative model, only at the level of the universal distribution (the distribution governing all computer programs), and (2) AID by way of CTM and BDM provides the means to implement other computational-mechanical approaches, including potentially Crutchfield's, as the means to mine the space of computable finite (deterministic or stochastic) automata.

\begin{figure}
\centerline{\includegraphics[scale=0.27]{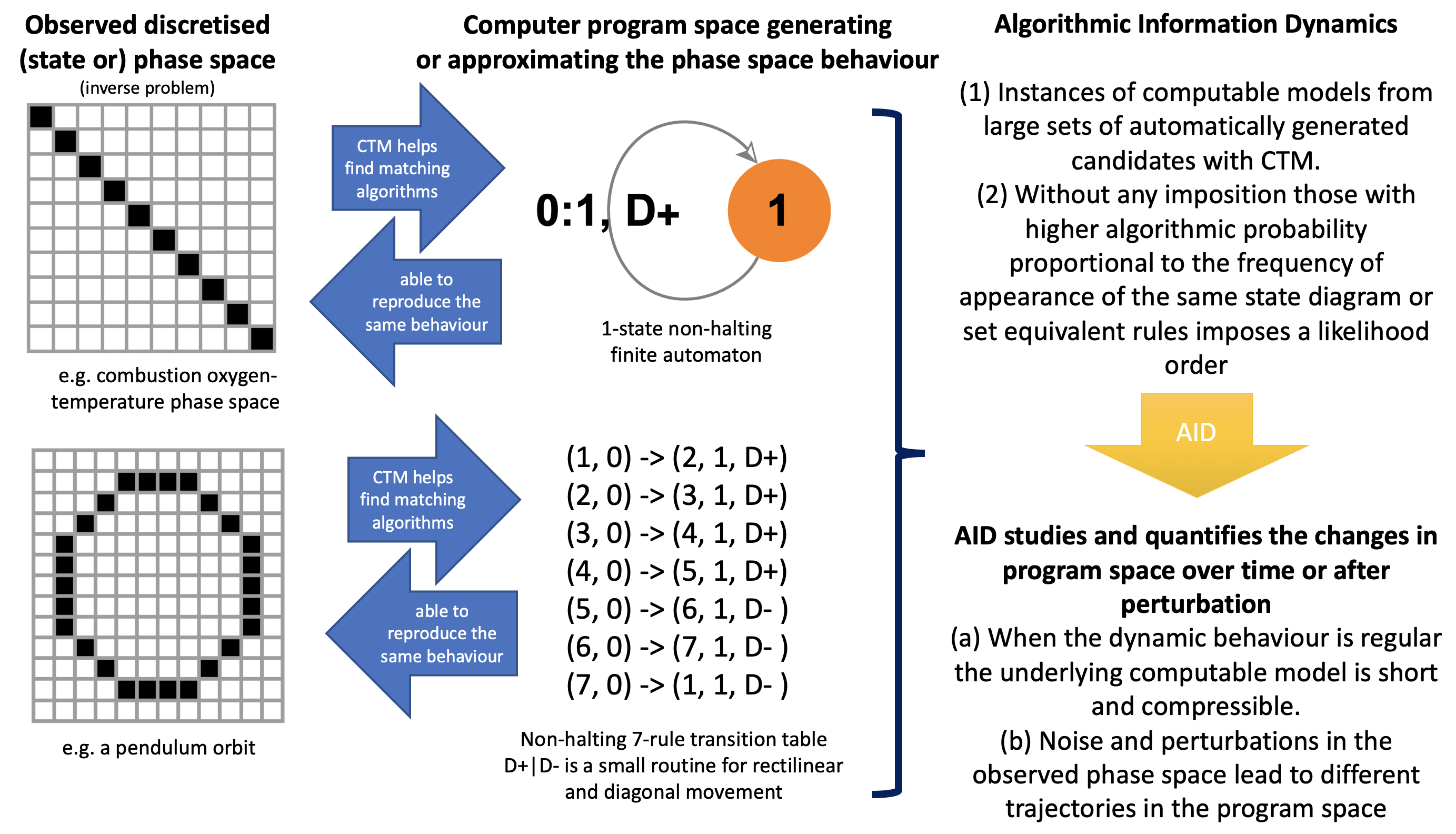}}
\caption{The basic principles of Algorithmic Information Dynamics consist in finding computable models or sequences of (short) computable models able to explain a piece of data or an observation and study the effect that interventions have on such models in the software space. Fully deterministic systems subject to no noise or no external influence will find an invariant set of candidate models with description lengths varying by only a logarithmic or sublogarithmic term; any deviation will suggest noise or external influence.\\}
\end{figure}

Because AID has the potential property (under assumptions of optimality) to distinguish simple from random and assign them similar values, AID can be considered a measure of sophistication similar to Bennett's logical depth. Moreover, a measure of (re)programmability is a measure of sophistication by design~\cite{zenilprogrammability}, and is based on AID. Some thermodynamic-like properties associated with reprogramming systems have been found and reported in the literature~\cite{zenil_2019b}.

Unlike other approaches, such as Pearl's do-calculus~\cite{pearl}, algorithmic information dynamics can help in the initial process of causal discovery and is able to perform fully unsupervised hypothesis generation from causal computable models. It also provides the tools to explore the algorithmic effects that perturbations to systems and data may have on their underlying computable models, effectively also providing a framework for causal analysis similar to the do-calculus, but without recourse to traditional probability distributions. AID can substitute for or complement other approaches to causal analysis. For example, the methods for causal analysis developed by Judea Pearl et al.~\cite{pearl} assume the existence of an educated causal model but cannot provide the means for primary causal discovery.

\begin{figure}
\centerline{\includegraphics[scale=0.8]{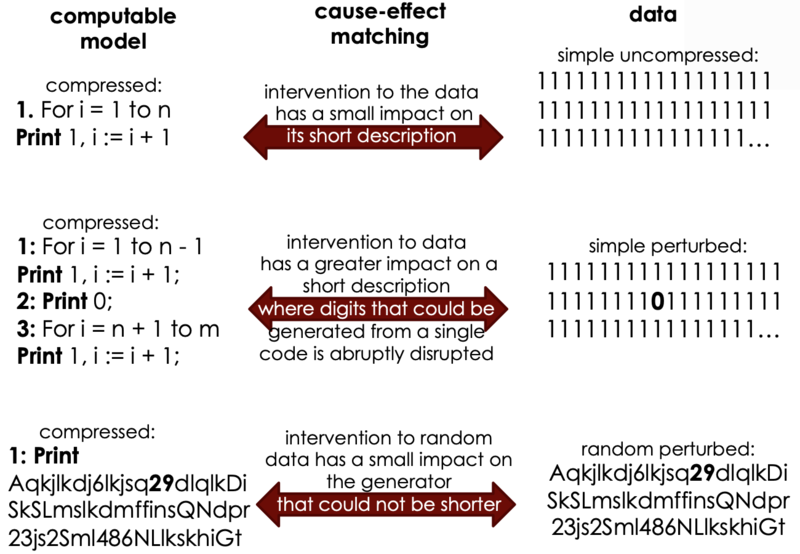}}
\vspace{0.5cm}
\caption{Causal intervention matching with AID: Central to AID is the study of the effect of interventions at the data level to the set of computable candidate models, their changes in program space being an indication both of the nature of the original data and the nature of the perturbation.\\}
\end{figure}

One can formulate a cause-effect question in the language of the do-calculus as a probability between a random variable and an intervention $P(L|do(D))$, and one can also do so in the context of AID. In one of the typical examples used by Pearl himself~\cite{pearl}, if $L$ represents the human lifespan and $do(D)$ the use of some drug $D$, the probability $P$ of $D$ having an effect on $L$ is quantified by classical probability to calculate $P$. What AID does is to substitute AP for $P$, the algorithmic probability that $D$ exerts an effect on $L$, with the chief advantage that not only does one obtain a probability for the dependency (and direction) between $D$ and $L$, but also a set of candidate models explaining such a connection, with no classical probability involved in the inference or description of each individual model. AP would then impose a natural non-uniform algorithmic probability distribution over the space of candidate models connecting $D$ and $L$ based on estimations of the universal distribution, which in effect introduces a simplicity bias that favors shorter models as opposed to algorithmically random ones, all of which have already been found to explain the causal connection between D and L. AID thus removes the need for classical probability distributions, and more importantly produces a set of generative models no longer derived from traditional statistics (e.g. regression, correlation), which even the do-calculus uses to determine the probability of a dependency, thereby falling back on the methods the do-calculus set out to circumvent.

While perturbation analysis is a major improvement in the area of causal discovery, AID complements it, offering a path to leave the use of classical probability descriptions out of the models taking a step further towards independence from other causal confounding statistical methods that introduce probability in the model description obfuscating first principles and preventing generative models. Another key difference between the original do-calculus and the algorithmic probability calculus is that the do-calculus makes a distinction between $P(L|do(D))$ and $P(L|D)$, but in AID both hypotheses $AP(L|D)$ and $AP(D|L)$ can be tested independently, as they have different meanings under algorithmic probability. $AP(L|D)$ means that $L$ is the generative mechanism of $D$ and $AP(D|L)$ means that $D$ is the generative mechanism of $L$. Each of them would trigger a different explorative process under AID. The former would look for the set of smaller to larger computable models denoted by L that can explain $D$, while the latter would look for all the generative models $\{D\}$ that generate $L$. Intuitively, this means that, for example, an attempt to explain how a barometer falling ($X$) could be the cause for a storm would likely not yield many shorter generative mechanisms to causally explain the occurrence of the storm ($Y$), whereas an attempt to do the reverse would, indicating that barometric fall is effect rather than cause. Clearly, regardless of the result, $AP(X|Y)$ and $AP(Y|X)$ are not equivalent.

In the language of AID, conforming more to a typical notation in graph and set theory, the first example would be written as $AP(L\backslash D)$ or $C(L\backslash D)$, where $C$ is the algorithmic complexity of L with intervention $D$ (in the context of AID this is usually a deletion to a model that includes D), and the second example as $AP(X\backslash Y)$ or $C(X\backslash Y)$. Alternatively, $AP(L\backslash {D})$ would be L with a set of interventions $\{D\}$. Multivariate causal analysis can be achieved by conditional versions of algorithmic complexity and is currently an area of active research.

In Pearl's account of causal analysis~\cite{pearl}, the so-called causal ladder leading up to human-grade reasoning consists of three levels, with the first being that of pattern observation covered by traditional statistics and long based on regression and correlation, the second being interventions such as the one his do-calculus suggests and that AID allows, and the 3rd level is that of counterfactuals or the power to imagine what would happen if conditions were different. AID also provides the most fundamental step in the causal analysis, which is causal discovery, within a single framework and without the need of resorting to making recourse to different approaches for different purposes (discovery vs analysis). While the do-calculus comes with no initial guidelines for how to come up with a first testable model, AID can generate a set of computable models ranked by likelihood and offers a path towards the full replacement of regression and correlation in the description of a model, taking the statistical nature out of the causal description. In addition, AID can also potentially cover all three levels of causality in Pearl's ladder, and provides the methodological framework to address each without the need of classical probability, regression, or correlation.


\subsection{Information Deficiency as an Algorithmic Information Calculus}

With its ability to quantify the departure of models away from or toward algorithmic randomness by perturbing a piece of data or a system, AID enables the investigation of which elements of the data contribute the most to the information necessary for the underlying causal computable model (the positive information elements) and which elements would trigger an increase in the system's algorithmic complexity (the negative information elements)~\cite{Zenil2019, zenil_2019b}. This way, the ``algorithmic randomness control" performed by the algorithmic intervention analysis puts forward new general methods for studying computable perturbation effects on non-linear systems, beyond those derived from classical control theory, and without making strong a priori assumptions of linearity. For example, AID has shown how to reconstruct the space-time evolution of Elementary Cellular Automata by gathering disordered states and rearranging by their constructive perturbation value in software space using AID~\cite{Zenil2019}.

Let $\left| S \right|$ denote the size of the object $S$, and not only the number of constitutive elements. Let $N$ denote the total number of constitutive elements of $S$. 


Let $F\neq \emptyset$ be a subset of the set of elements of the object $S$ to be computably perturbed. In order to grasp such a formal measure of the algorithmic-informational contribution of each element, one may look into the difference in algorithmic complexity between the original data $S$ and the perturbed data $S\backslash F$. We define the information difference ~\cite{Zenil2019, zenil_2019b} between $S$ and $S\backslash  F$ as,

\begin{equation}
I(S,F)=K(S)-K(S\backslash F).
\end{equation}

which is analog to randomness deficiency but between two  not necessarily random states.

\subsection{Study of Dynamical Systems in Software Space}

\begin{figure}[htbp]
    \centering
    \includegraphics[width=11cm]{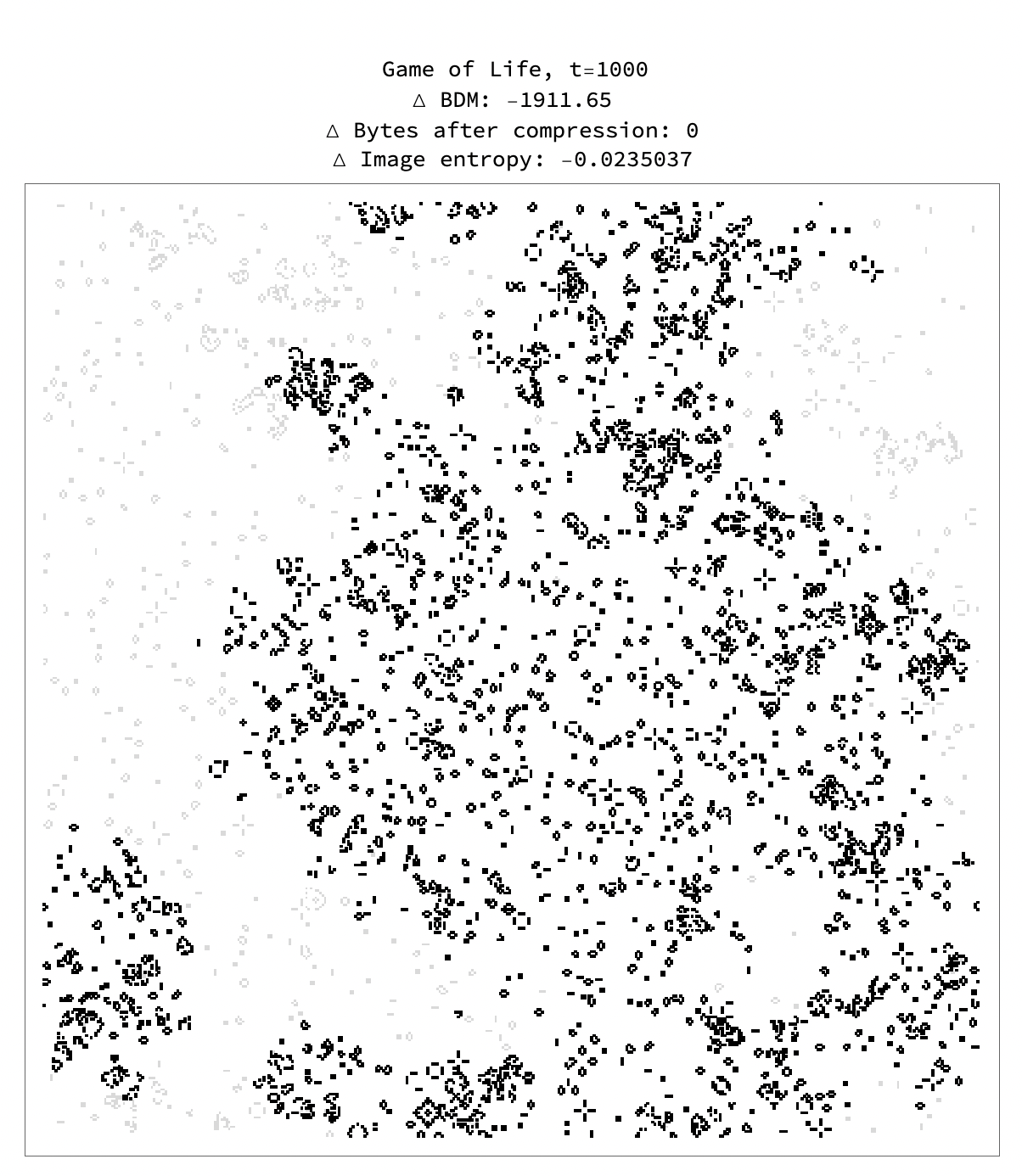}
    \caption{Perturbation of a single central cell in the 2D Cellular Automaton Game of Life found by John, H. Conway after running for 1000 steps from an initial random configuration producing its typical particle structures. The resulting BDM change from the perturbation is showed along with the change in bytes after compression, along with the entropy differences between the perturbed (black) and unperturbed cells (grey).\\}
    \label{gol}
\end{figure}

While some theoretical results have connected algorithmic complexity to dynamical systems, not many applications have done so. One of the main results in AID is that if a chain of causally connected processes is unaffected and generated from the same generative mechanism, then its algorithmic complexity remains constant up to a (double) logarithmic term accounting for the time step when the system is to be reproduced up to a certain observable time~\cite{Zenil2019}. Anything departing from such a quantity indicates that the process has been subject to an external perturbation or interaction which can be pinpointed by AID. Numerical experiments quantifying the change in the number of attractors in a dynamical system (and therefore the average depth or shallowness of these attractors) demonstrates that, on the one hand, removing elements that reduce the algorithmic complexity of a dynamical system (with respect to the average length of the models explaining its original description) systematically reduces the number of attractors ~\cite{Zenil2019}. On the other hand, when removing elements that increase the dynamical system's algorithmic complexity (thus making it more algorithmically random), the number of attractors increases. Current open topics of AID research include multivariable modeling, e.g., describing multiple particles individually rather than as a system.

The computational resources needed to compute CTM, at the core of AID, limit the size of its application, but BDM extends and combines CTM with other computable measures, including Shannon entropy itself. One of the most active areas of AID research is to find ways to make the algorithmic measures more relevant and to find possible shortcuts towards the computation of the computable region in the uncomputable space that is relevant for applications in science.

Similar methods based on AID have been proposed to equip machine intelligence with an inference engine based on AID, in order to help AI algorithms build computable hypotheses from data that can then be tested against observation~\cite{Hernandez-Orozco2018-hc}. This would complement other approaches such as statistical machine learning that fail at tasks such as inference and abstraction.

\section{AID Application to Elementary Cellular Automata}

We use BDM to approximate $-I(S,F)$ and define 

$$\Delta
BDM(S,F) = BDM(S\backslash F) - BDM(S),$$

\noindent where $S$ is the space-time evolution of an ECA and $S\backslash F$ is the same ECA after perturbation $F$ (e.g. a $n$ pixels in its initial condition). A type of algorithmic mutual information that can be interpreted as how much information a whole object has about the information of each of its constituent elements approached by, for example, perturbation analysis (perturbation or deletion of the elements of interest). To this end, we take several characteristic ECA examples belonging to different Wolfram classes.

\subsection{ECA Perturbation Analysis}

\begin{figure}[!h]
    \centering
    \small{LZW} \hspace{6cm} \small{BDM}\\
        \includegraphics[width=0.5\textwidth]{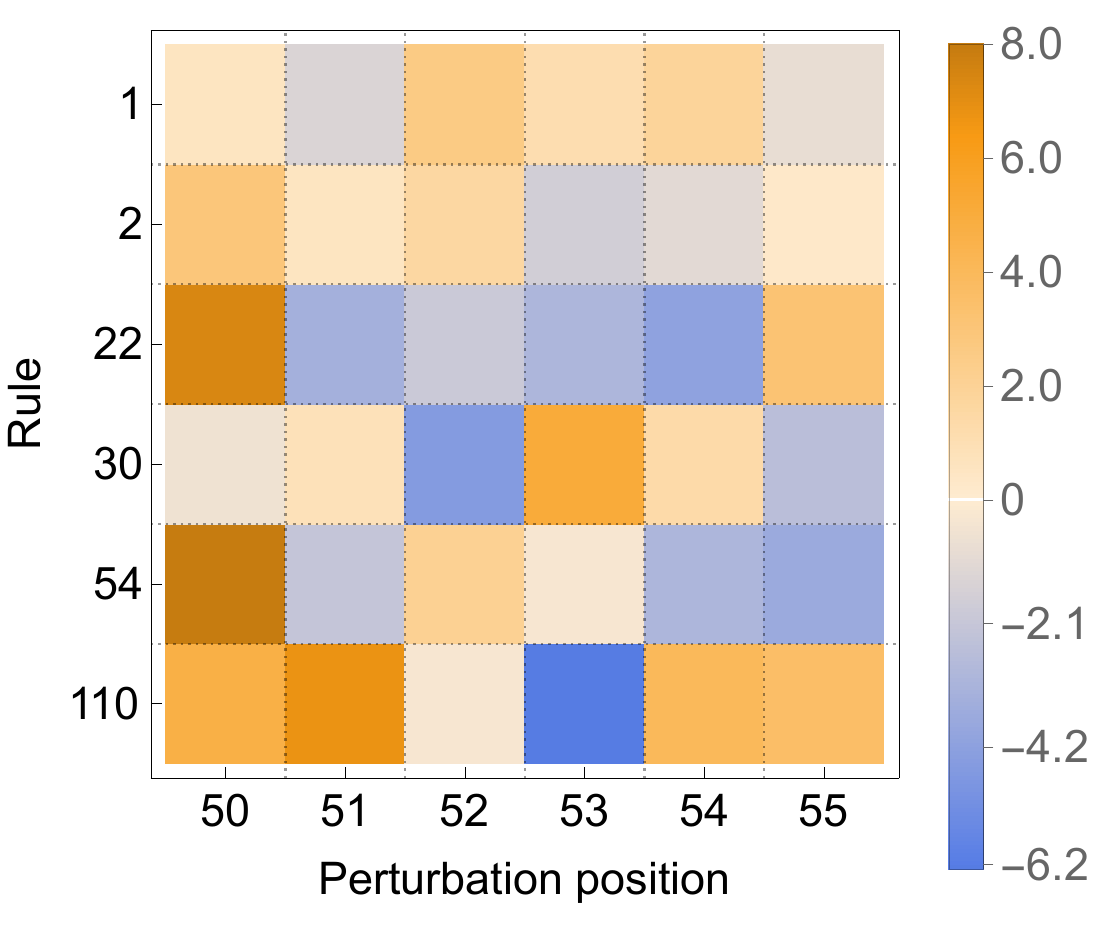}\includegraphics[width=0.5\textwidth]{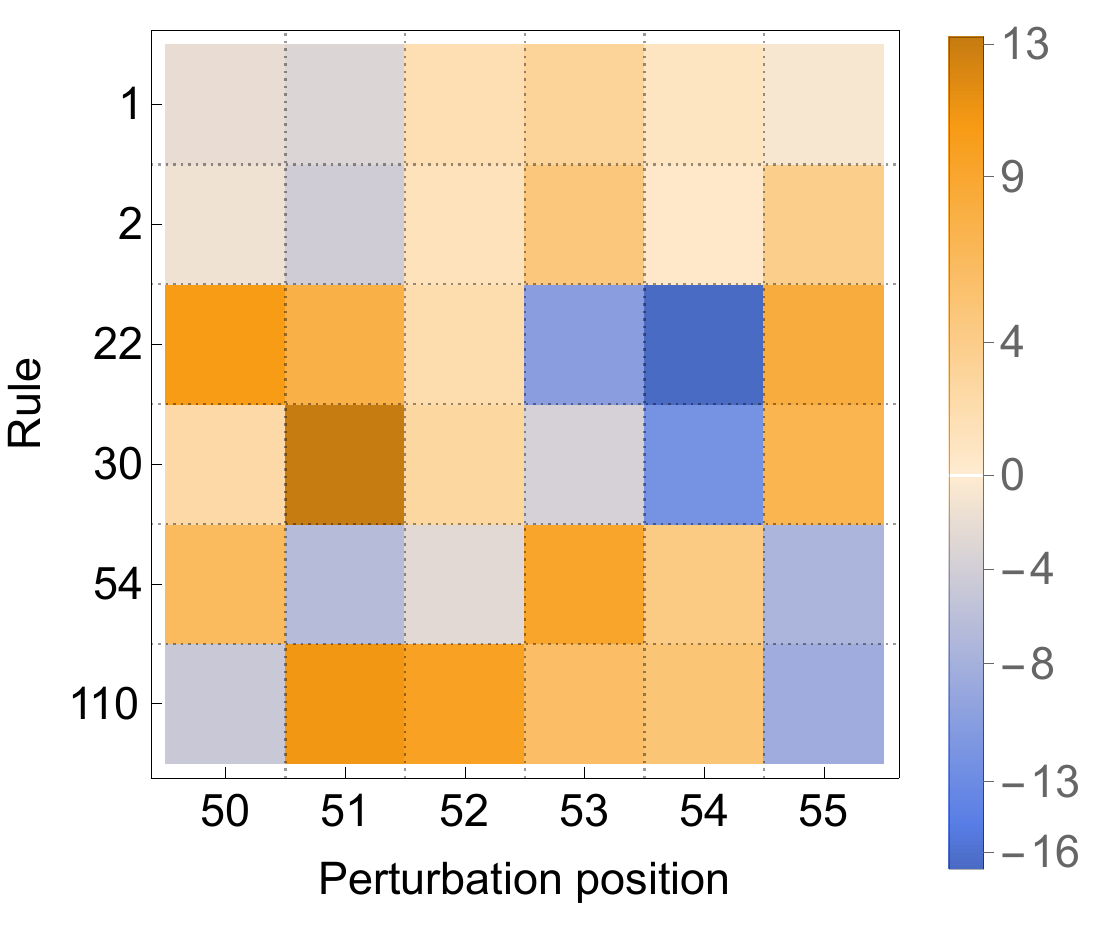}
    \caption{Left: Perturbation heatmap by compression (LZW). Colour indicates $\Delta$ number of bytes after compression over various perturbation positions on the initial state for 6 characteristic ECA rules. Right: Perturbation heatmap by BDM. Colour indicates $\Delta BDM$ over various perturbation positions on the initial state for six characteristic ECA rules. \\}
    \label{heatmaps1}
\end{figure}

As shown in \ref{heatmaps1}, BDM is more sensitive to small perturbations and provides finer grained and often divergent values to those found by statistical means (such as LZW).  The next experiment confirmed this and illustrates the way in which AID quantifies change not only in a precise manner but also with an underlying mechanistic explanation in the form of a computable model (in fact, a set of computable models sorted by likeliness according to algorithmic probability).

\begin{figure}[htbp]
    \centering
    \includegraphics[width=0.6\textwidth]{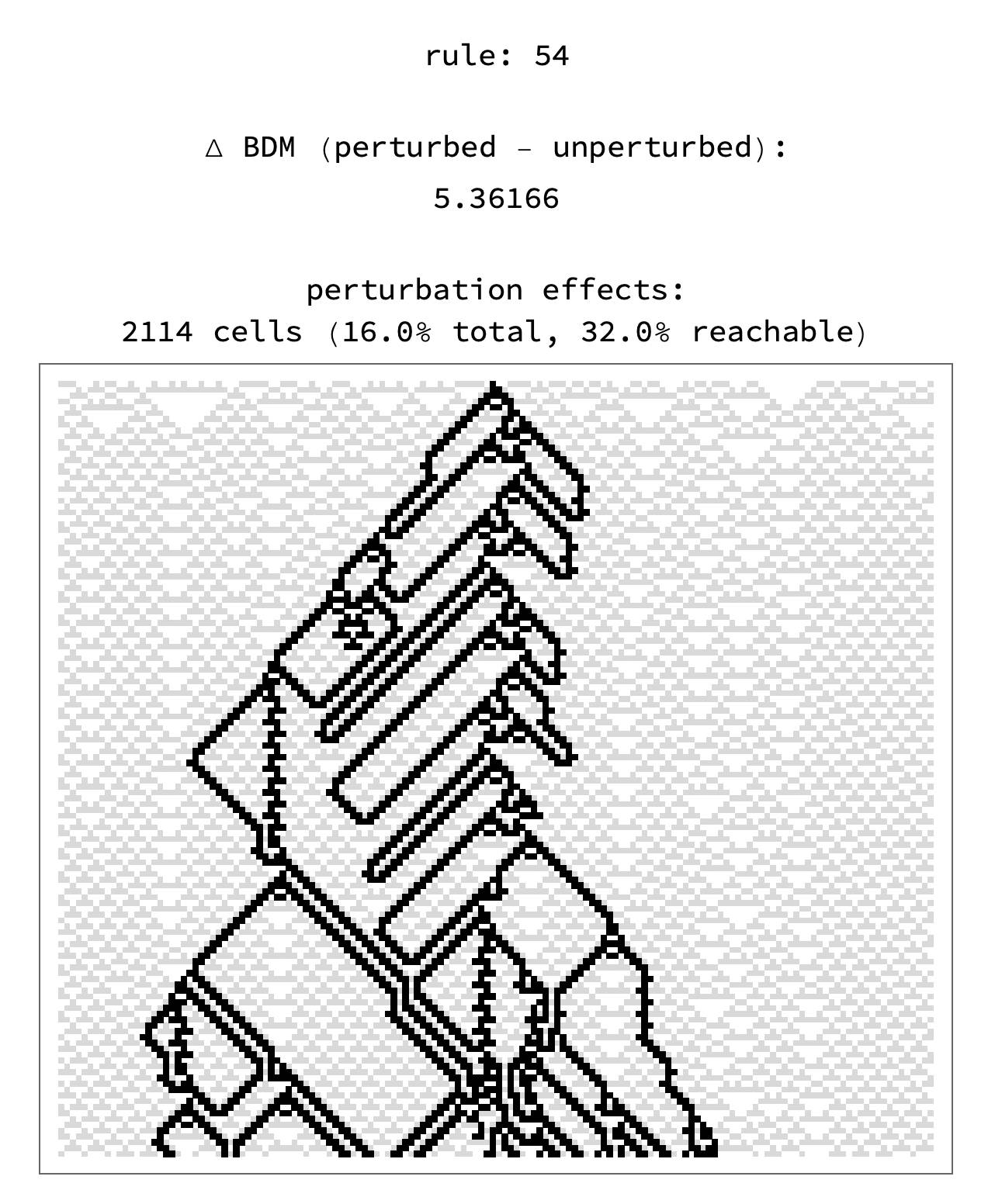}
    \caption{Sensitivity to small perturbations propagating over time. Single-cell perturbation on longer evolutions of Rule 54 with larger initial states. The perturbation analysis suggested by AID and implemented with BDM is sensitive enough to pick up the changes despite the apparent random background. The resulting value suggests the appearance of a second order flow of information (on a particle-like background already linearly transferring information) after a perturbation propagating in time towards the output effectively implementing a logical circuit.\\}
    \label{54examples1}
\end{figure}

\begin{figure}
    \centering
    \includegraphics[width=0.6\textwidth]{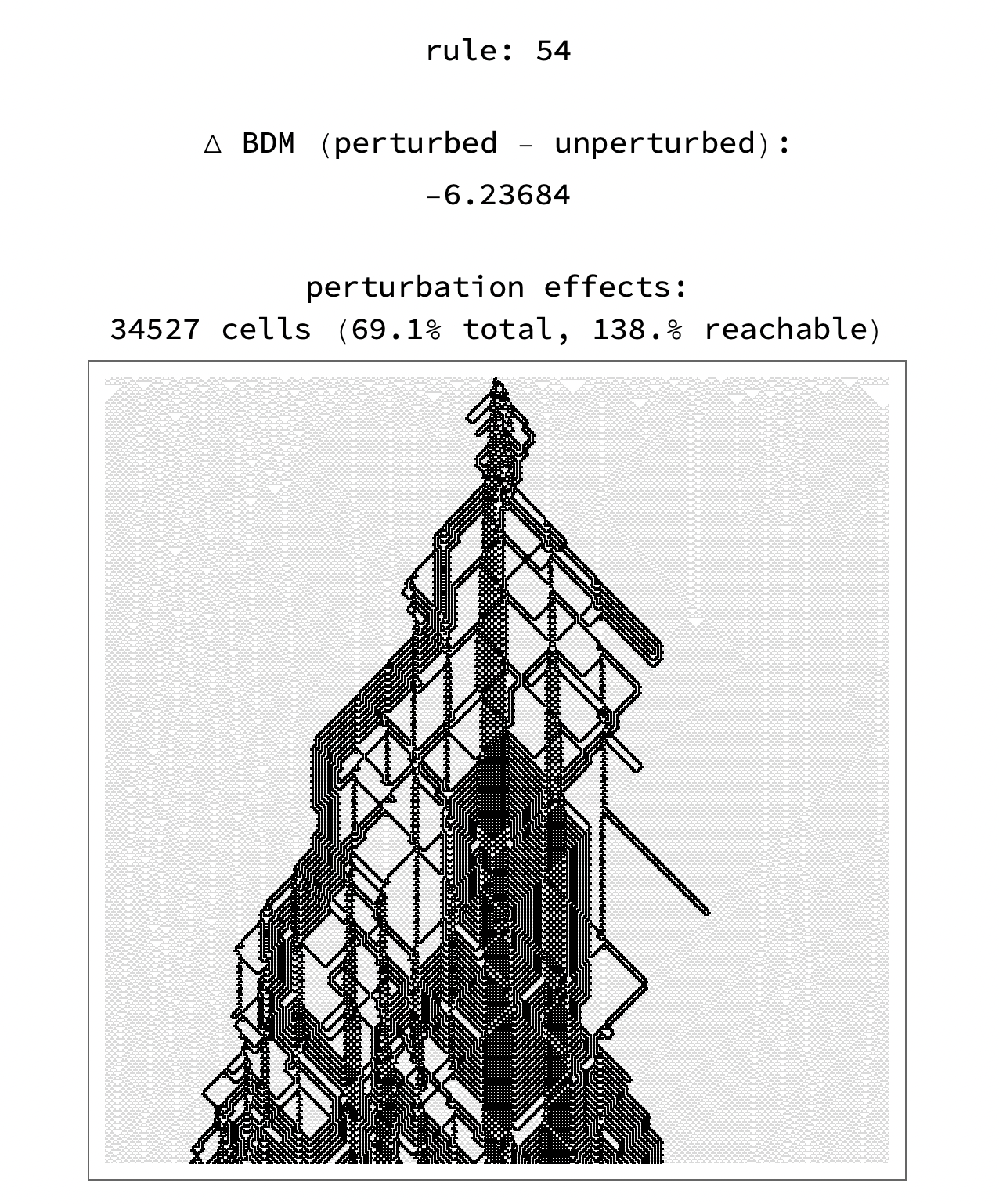}
    \caption{A longer runtime (and different random initial condition) of the perturbation yields interesting patterns with the perturbation propagating at different submaximal speeds on each side and flipping the sign of the BDM effectively indicating a change of algorithmic information dynamics from an increasing orderly circuit-like pattern to a random-looking pattern after multiple collisions from a perturbation that takes over the original background.\\}
    \label{54examples3}
\end{figure}

To test some of these ideas, we measured the differences in complexity (via BDM) of ECA evolutions with a single-cell perturbation in the initial state. For simplicity, we selected a small subset of representative ECA rules (1, 2, 22, 30, 54, and 100) spanning all Wolfram qualitative behavioural classes to run over random initial states for some amount of time. All random initial states were 100 cells long, and with periodic boundary conditions, each space-time `image' was generated by evolving the initial state over a rule for 80 time steps. 

To perturb each space-time evolution, a single cell in the initial state was selected and ``flipped'' at a cell position, where a 0 cell would become a 1 and vice versa. The perturbed state was evolved in the same way as the unperturbed state.

The BDM of each perturbed and unperturbed space-time evolution $C$ were compared as a change in $\Delta BDM = BDM(S_{perturbed}) - BDM(S_{unperturbed}$). We also compared this change in BDM with the measured the change in the number of bytes after space-time compression between the two evolutions. Figures \ref{heatmaps1} show the resulting change in BDM over various ECA rules and perturbation locations. Since the initial states are random, these rules show the average results over 1000 different random initial states.

Figures \ref{54examples1}, and \ref{54examples3} shows some examples of a single-cell perturbation on a rule 54 ECA with larger initial state sizes and longer time evolutions.


\subsection{ECA Colliding Event Quantification}

In addition to single-celled perturbations, we also measured the complexity of ECA that collide spatially. Each ECA instance shares the same set of 0 cells, but one ECA operates with 1 cells and the other operates with $-1$ cells. Because ECA evolve in a state space consisting of only two types of cells, the interaction rule must accommodate outcomes for three cell types. This interaction rule that determined the outcome for a neighborhood of three different cell types was chosen randomly as it is not important for the proof of concept, the quantification of the emergent structures as a result of the collision as an example of interacting dynamical systems. Since there are several different types of outcomes that a neighborhood of 3 different cell types could yield, 1000 different rule outcomes were sampled. The interaction rule preserves the individual ECA rule outcomes for cells whose neighborhoods consist of two cell types.

\begin{figure}[htbp]
    \centering
    \includegraphics[width=9.5cm]{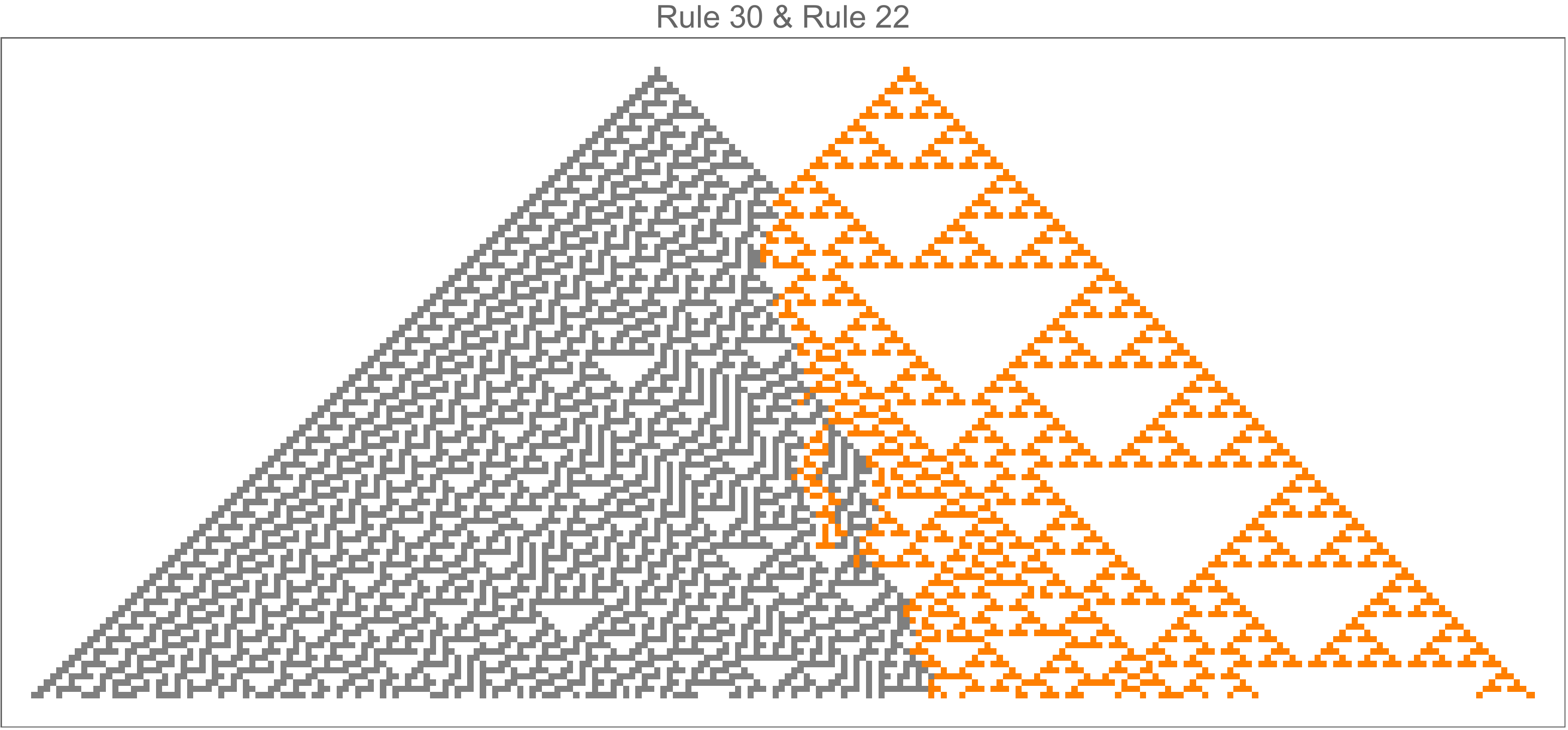}\\
    \includegraphics[width=9.5cm]{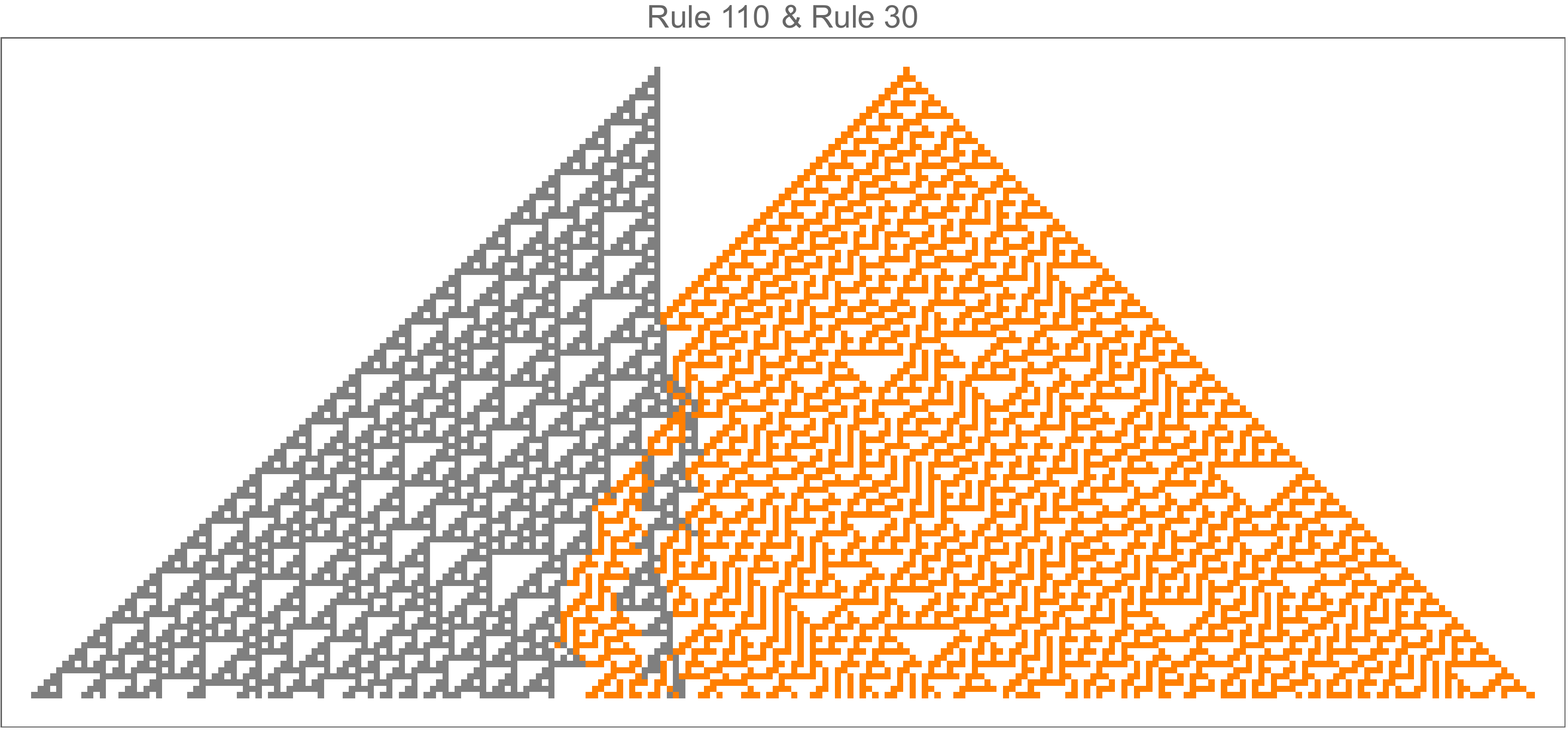}
    \caption{Lateral collision examples. Time evolutions of ECA rules 30 (grey) and 22 (orange) and rules 110 (grey) and 30 (orange) on simplest initial state (single black cell) in a collision under an arbitrary interaction rule.\\}
    \label{coll1}
\end{figure}

\begin{figure}[htbp]
    \centering
    \includegraphics[width=0.49\textwidth]{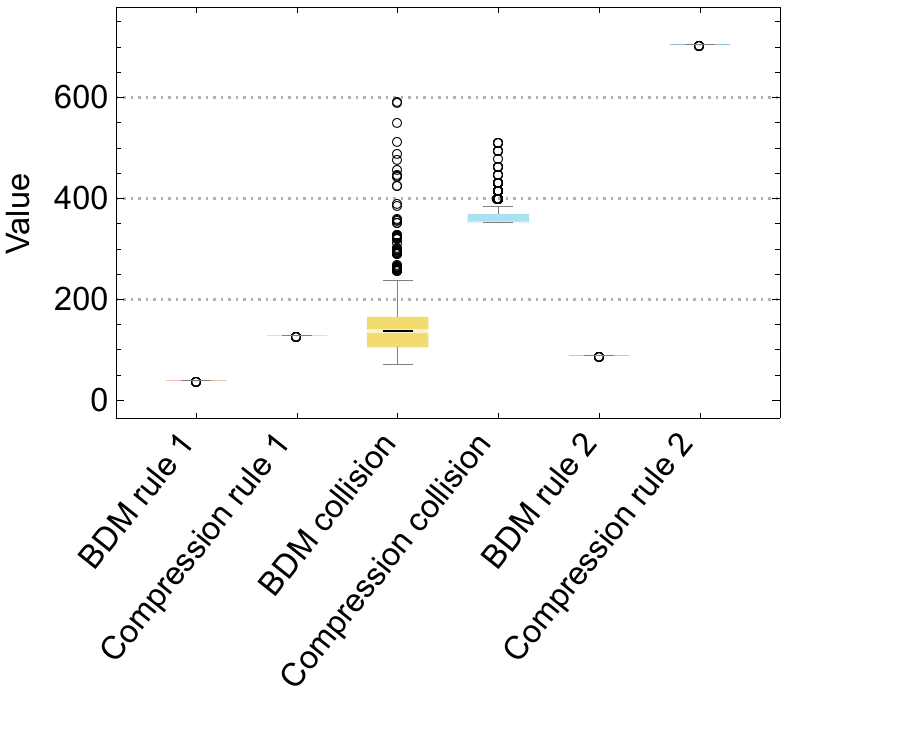} \includegraphics[width=0.49\textwidth]{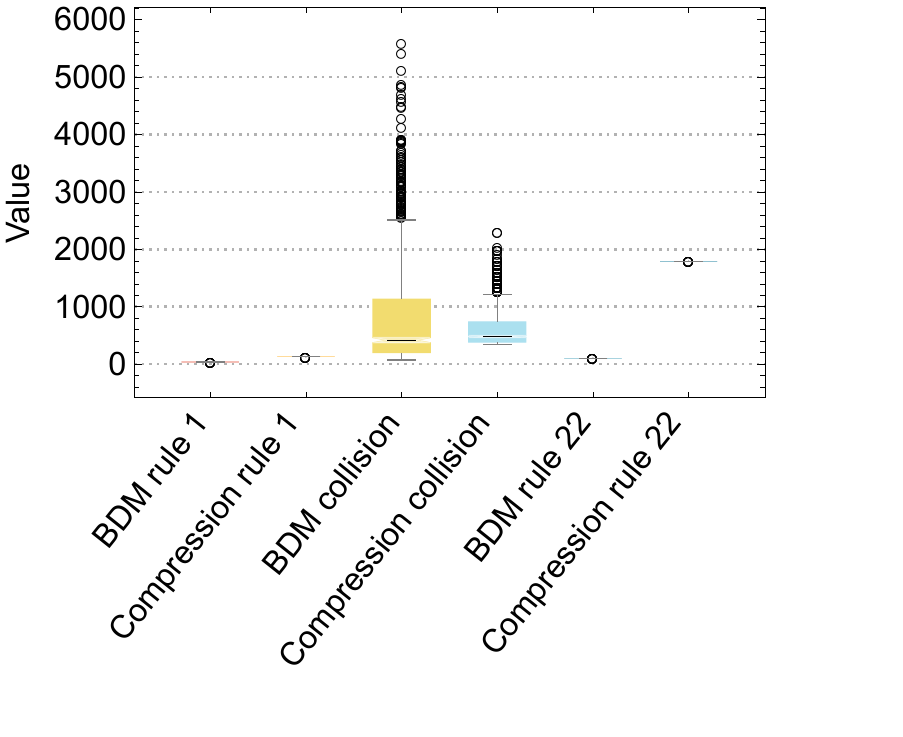}\\
    \includegraphics[width=0.49\textwidth]{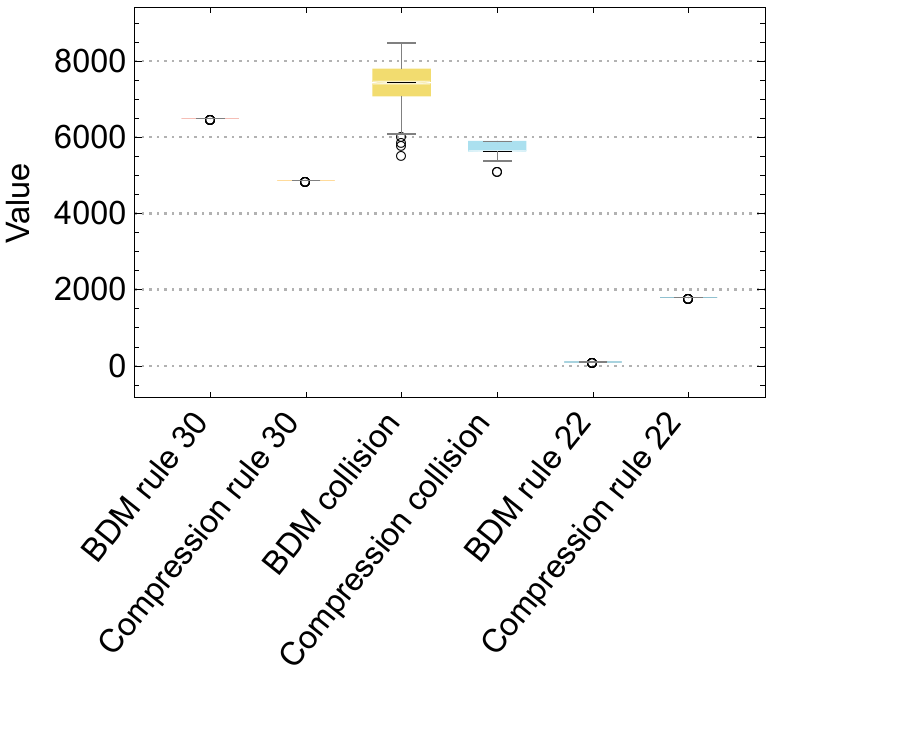} \includegraphics[width=0.49\textwidth]{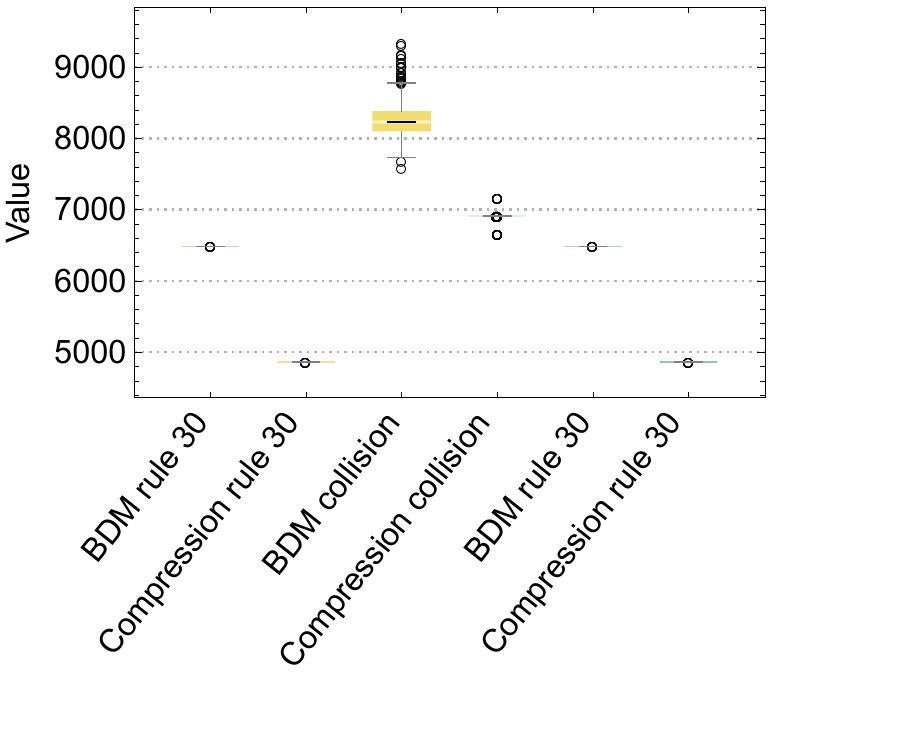}
    \caption{Top: Elementary CA (ECA) collisions Left: rule 1 v 2. Right: rule 1 v 22. Bottom: Wolfram Class 3 ECA collisions. Left: rule 30 v 22. Right: rule 30 v 30. Box plots show differences in BDM and compression during a CA collision, compared to these values in an isolated version of the ECA rule under a simple (single black cell) initial state.\\}
    \label{simple1}
\end{figure}

Each ECA were seeded as a single 1 and $-1$ with 40 cells of state 0 in-between. This initial state was evolved for 100 time steps under this interaction rule. The BDM of the resulting image was measured and compared to the BDM of each ECA image in isolation, without the interaction. The same was measured for the number of bytes after compression. Figures \ref{simple1}, show the differences in BDM and compression between ECA in isolation and during a collision event, over 1000 different interaction rules and for various ECA rule interaction types, by complexity class. For illustrative purposes, some example state evolutions are shown in Figures \ref{coll1}.

\section{Conclusions}

AID grants access to tools able to provide insights into first principles and mechanistic explanations. It is a tool based on the same kind of simple programs John Conway thought could help characterise other programs like the Game of Life as an oversimplification of life, for which connections to algorithmic complexity have also been established~\cite{zenillife}.

We have here shown that BDM powered by CTM allows the study of dynamical systems in software space, that is, the effect of a perturbation on the size of the set of underlying computable candidate models constructively generating each output.  In particular, we have demonstrated an application of AID to (elementary and 2D) cellular automata. We have shown how AID can be used to study discrete dynamical systems such as cellular automata and their evolution as an illustration of what AID can contribute to the study of dynamical systems from a computable-model perspective (what we call software space).

Based on these results, BDM was more likely to capture small changes from a perturbation than other measures, such as statistical (LZW) compression.  As illustrated in Figure~\ref{gol}, typical statistical measures such as Shannon entropy or LZW and cognates will be insensitive to small changes (because of lack of granularity, its shortcomings inherited from Shannon entropy, and operating check-sums) specially on a disordered background, failing at enabling the principles of algorithmic information dynamics.

\subsection*{Acknowledgements}

We wish to thank Abicumaran Uthamacumaran for his help with preparing some parts of this document in the required format.

\printbibliography

@ARTICLE{zenilreview,
AUTHOR = {Zenil, H.},
TITLE   = {A Review of Methods for Estimating Algorithmic Complexity: Options, Challenges, and New Directions},
YEAR    = {2020},
JOURNAL = {Entropy},
VOLUME  = {612},
NUMBER  = {22}
}

@ARTICLE{Zenil:2020,
AUTHOR = {Zenil, Hector  and Kiani, Narsis A. and Abrahão, Felipe S and Tegnér, Jesper N.},
TITLE   = {{A}lgorithmic {I}nformation {D}ynamics},
YEAR    = {2020},
JOURNAL = {Scholarpedia},
VOLUME  = {15},
NUMBER  = {7},
PAGES   = {53143},
DOI     = {10.4249/scholarpedia.53143},
NOTE    = {revision \#195807}
}

@ARTICLE{conway,
AUTHOR = {Gardner, Martin},
TITLE   = {Mathematical Games: The fantastic combinations of John Conway's new solitaire game `Life'},
YEAR    = {1970},
JOURNAL = {Scientific American},
VOLUME  = {223},
PAGES   = {120–123}
}

@INCOLLECTION{zenilgol,
  title     = "Algorithmic Information Dynamics of Emergent, Persistent, and Colliding Particles in the Game of Life",
  booktitle = "From Parallel to Emergent Computing",
  author    = "Hector Zenil and Narsis A. Kiani and Jesper Tegnér",
  publisher = "Taylor \& Francis / CRC Press",
  pages     = "367--383",
  year      =  2019
}

@BOOK{Calude2002-da,
  title     = "Information and Randomness",
  author    = "Calude, Cristian S",
  publisher = "Springer",
  series    = "Texts in Theoretical Computer Science. An EATCS Series",
  month     =  dec,
  year      =  2002,
  address   = "Berlin, Germany",
  language  = "en"
}

@BOOK{Chaitin1987-kt,
  title     = "Algorithmic Information Theory",
  author    = "Chaitin, Gregory J",
  publisher = "Cambridge University Press",
  month     =  oct,
  year      =  1987
}

@BOOK{nks,
  title     = "A New Kind of Science",
  author    = "Stephen Wolfram",
  publisher = "Wolfram Media",
  year      =  2002,
  address   = "Champaign, IL"
}

@BOOK{pearl,
  title     = "CAUSALITY: Models, Reasoning, and Inference",
  author    = "Judea Pearl",
  publisher = "Cambridge University Press",
  year      =  2000,
  address   = "Cambridge, UK"
}

@ARTICLE{StephenWolfram1983,
  title   = "Statistical Mechanics of Cellular Automata",
  author  = "Stephen Wolfram",
  journal = "Reviews of Modern Physics",
  volume  =  55,
  number = 3,
  pages   = "601–644",
  year    =  1983
}

@ARTICLE{zenilchaos,
  title   = "Asymptotic Behaviour and Ratios of Complexity in Cellular Automata Rule Spaces",
  author  = "Hector Zenil",
  journal = "International Journal of Bifurcation and Chaos",
  volume  =  23,
  number = 9,
  year    =  2013
}

@ARTICLE{zenilgeo,
  title   = "Symmetry and Correspondence of Algorithmic Complexity over Geometric, Spatial and Topological Representations",
  author  = "Hector Zenil and Narsis A. Kiani and Jesper Tegnér",
  journal = "Entropy",
  volume  =  20,
  number = 7,
  issue = 534,
  year    =  2018
}

@INCOLLECTION{Delahaye2007-oz,
  title     = "On the kolmogorov-Chaitin complexity for short sequences",
  booktitle = "Randomness and Complexity, From Leibniz to Chaitin",
  author    = "Delahaye, Jean-Paul and Zenil, Hector",
  publisher = "World Scientific Publishing Press Publishing Press",
  pages     = "123--129",
  month     =  oct,
  year      =  2007
}

@ARTICLE{Delahaye2012-eb,
  title     = "Numerical evaluation of algorithmic complexity for short
               strings: A glance into the innermost structure of randomness",
  author    = "Delahaye, Jean-Paul and Zenil, Hector",
  journal   = "Appl. Math. Comput.",
  publisher = "Elsevier BV",
  volume    =  219,
  number    =  1,
  pages     = "63--77",
  month     =  sep,
  year      =  2012,
  language  = "en"
}

@BOOK{Downey2010-tq,
  title     = "Algorithmic randomness and complexity",
  author    = "Downey, Rodney G and Hirschfeldt, Denis R",
  publisher = "Springer New York",
  series    = "Theory and applications of computability",
  year      =  2010,
  address   = "New York, NY"
}

@ARTICLE{zenillife,
  title   = "Life as Thermodynamic Evidence of Algorithmic Structure in Natural Environments",
  author  = "Hector Zenil, and Carlos Gershenson, and James A.R. Marshall and David Rosenblueth",
  journal = "Entropy",
  volume  =  14,
  number = 11,
  pages   = "2173--2191",
  year    =  2012
}

@ARTICLE{Hernandez-Orozco2018-hc,
  title     = "Algorithmically probable mutations reproduce aspects of
               evolution, such as convergence rate, genetic memory and
               modularity",
  author    = "Hern{\'a}ndez-Orozco, Santiago and Kiani, Narsis A and Zenil,
               Hector",
  journal   = "R. Soc. Open Sci.",
  publisher = "The Royal Society",
  volume    =  5,
  number    =  8,
  pages     = "180399",
  month     =  aug,
  year      =  2018,
  language  = "en"
}

@ARTICLE{Zenil2019,
  title     = "An algorithmic information calculus for causal discovery and
               reprogramming systems",
  author    = "Zenil, Hector and Kiani, Narsis A and Marabita, Francesco and
               Deng, Yue and Elias, Szabolcs and Schmidt, Angelika and Ball,
               Gordon and Tegn{\'e}r, Jesper",
  journal   = "iScience",
  publisher = "Elsevier BV",
  volume    =  19,
  pages     = "1160--1172",
  month     =  sep,
  year      =  2019,
  copyright = "http://creativecommons.org/licenses/by-nc-nd/4.0/",
  language  = "en"
}

@article{zenil_2018, title={A Decomposition Method for Global Evaluation of Shannon Entropy and Local Estimations of Algorithmic Complexity}, volume={20}, DOI={10.3390/e20080605}, number={8}, journal={Entropy}, author={Zenil, Hector and Hernández-Orozco, Santiago and Kiani, Narsis and Soler-Toscano, Fernando and Rueda-Toicen, Antonio and Tegnér, Jesper}, year={2018}, pages={605}}

@ARTICLE{zenilprogrammability,
AUTHOR = {Zenil, H.},
TITLE   = {What is nature-like computation? A behavioural approach and a notion of programmability},
YEAR    = {2014},
JOURNAL = {Philosophy \& Technology},
VOLUME  = {27},
PAGES = {399-421},
NUMBER  = {3}
}

@article{zenil_2019b, title={The Thermodynamics of Network Coding, and an Algorithmic Refinement of the Principle of Maximum Entropy}, volume={21}, DOI={10.3390/e21060560}, number={6}, journal={Entropy}, author={Zenil, Hector and Kiani, Narsis A. and Tegnér, Jesper}, year={2019}, pages={560}}

@BOOK{Li2019-op,
  title     = "An introduction to kolmogorov complexity and its applications",
  author    = "Li, Ming and Vit{\'a}nyi, Paul",
  publisher = "Springer International Publishing",
  series    = "Texts in computer science",
  year      =  2019,
  address   = "Cham"
}

@ARTICLE{Soler-Toscano2013-fm,
  title     = "Correspondence and independence of numerical evaluations of
               algorithmic information measures",
  author    = "Soler-Toscano, Fernando and Zenil, Hector and Delahaye,
               Jean-Paul and Gauvrit, Nicolas",
  journal   = "Computability",
  publisher = "IOS Press",
  volume    =  2,
  number    =  2,
  pages     = "125--140",
  year      =  2013
}

@INCOLLECTION{Zenil2020a,
  title     = "Compression is comprehension and the unreasonable effectiveness
               of digital computation in the natural world",
  booktitle = "Unravelling Complexity",
  author    = "Zenil, Hector",
  publisher = "World Scientific Publishing Press",
  pages     = "201--238",
  month     =  feb,
  year      =  2020
}

@ARTICLE{Soler-Toscano2014-er,
  title     = "Calculating kolmogorov complexity from the output frequency
               distributions of small Turing machines",
  author    = "Soler-Toscano, Fernando and Zenil, Hector and Delahaye,
               Jean-Paul and Gauvrit, Nicolas",
  journal   = "PLoS One",
  publisher = "Public Library of Science (PLoS)",
  volume    =  9,
  number    =  5,
  pages     = "e96223",
  month     =  may,
  year      =  2014,
  language  = "en"
}

@ARTICLE{Soler-Toscano2014-bj,
  title     = "Calculating kolmogorov complexity from the output frequency
               distributions of small Turing machines",
  author    = "Soler-Toscano, Fernando and Zenil, Hector and Delahaye,
               Jean-Paul and Gauvrit, Nicolas",
  journal   = "PLoS One",
  publisher = "Public Library of Science (PLoS)",
  volume    =  9,
  number    =  5,
  pages     = "e96223",
  month     =  may,
  year      =  2014,
  language  = "en"
}

@ARTICLE{Solomonoff1964-hm,
  title     = "A formal theory of inductive inference. Part {I}",
  author    = "Solomonoff, Ray J.",
  journal   = "Inf. Contr.",
  publisher = "Elsevier BV",
  volume    =  7,
  number    =  1,
  pages     = "1--22",
  month     =  mar,
  year      =  1964,
  copyright = "https://www.elsevier.com/open-access/userlicense/1.0/",
  language  = "en"
}

@ARTICLE{Zenil2015-dx,
  title     = "Algorithmicity and programmability in natural computing with the
               Game of Life as in silico case study",
  author    = "Zenil, Hector",
  journal   = "J. Exp. Theor. Artif. Intell.",
  publisher = "Informa UK Limited",
  volume    =  27,
  number    =  1,
  pages     = "109--121",
  month     =  jan,
  year      =  2015,
  language  = "en"
}

@ARTICLE{Levin1974-ci,
  title   = "Laws of Information Conservation (Nongrowth) and Aspects of the
             Foundation of Probability Theory",
  author  = "Levin, Leonid A",
  journal = "Problems Inform. Transmission",
  volume  =  10,
  number  =  3,
  pages   = "206--210",
  year    =  1974
}

@ARTICLE{Zenil2017a,
  title     = "Low-algorithmic-complexity entropy-deceiving graphs",
  author    = "Zenil, Hector and Kiani, Narsis A and Tegn{\'e}r, Jesper",
  journal   = "Phys. Rev. E.",
  publisher = "American Physical Society (APS)",
  volume    =  96,
  number    =  1,
  month     =  jul,
  year      =  2017,
  copyright = "https://creativecommons.org/licenses/by/4.0/",
  language  = "en"
}

@ARTICLE{Zenil2019-tk,
  title     = "Coding-theorem like behaviour and emergence of the universal
               distribution from resource-bounded algorithmic probability",
  author    = "Zenil, Hector and Badillo, Liliana and Hern{\'a}ndez-Orozco,
               Santiago and Hern{\'a}ndez-Quiroz, Francisco",
  journal   = "Int. J. Parallel Emergent Distrib. Syst.",
  publisher = "Informa UK Limited",
  volume    =  34,
  number    =  2,
  pages     = "161--180",
  month     =  mar,
  year      =  2019,
  language  = "en"
}

@INCOLLECTION{Zenil2020-xw,
  title     = "Compression is comprehension and the unreasonable effectiveness
               of digital computation in the natural world",
  booktitle = "Unravelling Complexity",
  author    = "Zenil, Hector",
  publisher = "World Scientific Publishing Press",
  pages     = "201--238",
  month     =  feb,
  year      =  2020
}








\end{document}